\documentclass[aps,prb,twocolumn,superscriptaddress]{revtex4-1}

\usepackage{graphicx}
\usepackage[ansinew]{inputenc}
\usepackage{amsmath,amsfonts,amssymb}

\begin{document}

\title{Electron and Cooper pair transport across a single magnetic molecule explored with a scanning tunneling microscope}

\author{J.\ Brand}
\affiliation{Institut für Physik, Technische Universität Ilmenau, D-98693 Ilmenau, Germany}
\author{S.\ Gozdzik}
\affiliation{Institut für Physik, Technische Universität Ilmenau, D-98693 Ilmenau, Germany}
\author{N.\ Néel}
\affiliation{Institut für Physik, Technische Universität Ilmenau, D-98693 Ilmenau, Germany}
\author{J.\ L.\ Lado}
\affiliation{Institute for Theoretical Physics, ETH Zürich, CH-8093 Zürich, Switzerland}
\affiliation{QuantaLab, International Iberian Nanotechnology Laboratory (INL), Avenida Mestre José Veiga, 4715-310 Braga, Portugal}
\author{J.\ Fernández-Rossier}
\altaffiliation{On leave from Departamento de Fisica Aplicada, Universidad de Alicante, Spain}
\affiliation{QuantaLab, International Iberian Nanotechnology Laboratory (INL), Avenida Mestre José Veiga, 4715-310 Braga, Portugal}
\author{J.\ Kröger}
\email{joerg.kroeger@tu-ilmenau.de}
\affiliation{Institut für Physik, Technische Universität Ilmenau, D-98693 Ilmenau, Germany}

\begin{abstract}
A scanning tunneling microscope is used to explore the evolution of electron and Cooper pair transport across single Mn-phthalocyanine molecules adsorbed on Pb(111) from tunneling to contact ranges.
Normal-metal as well as superconducting tips give rise to a gradual transition of the Bardeen-Cooper-Schrieffer energy gap in the tunneling range into a zero-energy resonance close to and at contact.
Supporting transport calculations show that in the normal-metal -- superconductor junctions this resonance reflects the merging of in-gap Yu-Shiba-Rusinov states as well as the onset of Andreev reflection.
For the superconductor -- superconductor contacts the zero-energy resonance is rationalized in terms of a finite Josephson current that is carried by phase-dependent Andreev and Yu-Shiba-Rusinov levels.
\end{abstract}

\maketitle

\section{Introduction}

The interplay between superconductivity and magnetism has been a longstanding topic of research.
It has recently witnessed a remarkable revival owing to investigations into Majorana zero modes and topological superconductivity induced by magnetic nanostructures on surfaces of $s$-wave superconductors. \cite{science_346_602,natphys_13_286,prl_115_197204,natcommun_7_12297,natcommun_8_2040}
On general grounds, it was predicted that spin-dependent scattering of Bogoliubov quasiparticles from magnetic impurities in conventional superconductors cause Yu-Shiba-Rusinov (YSR) states \cite{aps_21_75,ptp_40_435,jetpl_9_85} that modify the density of states (DOS) of the superconductor and appear as characteristic spectroscopic features within the Bardeen-Cooper-Schrieffer (BCS) \cite{pr_108_1175} energy gap.
Early experiments using planar proximity tunneling junctions of CuX (X$=$Mn, Cr, Fe) and AuFe alloys, \cite{prb_16_1068} thermal conductivity \cite{prb_14_1039} and specific-heat \cite{prb_16_1086} measurements of InMn, InCr, PbMn alloys as well as Mn-implanted Pb films \cite{prl_47_1163} were interpreted on the basis of YSR states.

Locally, \textit{i.\,e.}, at the single-atom level, the spectroscopic signature of YSR states was first reported in experiments with a scanning tunneling microscope (STM) on single Mn and Gd adsorbed on Nb(110) surfaces. \cite{science_275_1767}
In the spectra of the differential conductance ($\text{d}I/\text{d}V$), a pair of peaks appeared at sample voltages $\pm\varepsilon_{\text{B}}/\text{e}$ ($\varepsilon_{\text{B}}$: binding energy of the YSR state, e: elementary charge), reflecting the electron ($+$) and hole ($-$) component of the excitation.
The peak heights of these signatures were asymmetric, which was ascribed to the lack of electron-hole symmetry of the band structure \cite{prb_56_11213,prl_78_3761} or the nonmagnetic scattering potential of the impurity. \cite{rmp_78_373,prb_55_12648}

Renewed interest in YSR states has recently been initiated in $\text{d}I/\text{d}V$ spectroscopy of Mn and Cr atoms on superconducting Pb films on Si(111)\@. \cite{prl_100_226801,apl_96_073113}
In these reports, YSR states with different angular momentum could be resolved owing to the enhanced energy resolution that was achieved by low temperatures and the use of superconducting tips.
Moreover, in the seminal experimental work on the competition between Cooper pairing and Kondo screening \cite{science_332_940} it was further demonstrated that an increased exchange coupling between the magnetic impurity and the superconductor leads to a decrease of $|\varepsilon_\text{B}|$ and a crossing of $\varepsilon_\text{B}$ with the Fermi level accompanied by the inversion of the electron-hole spectral weight. 
Several other works followed presenting the requirement of topological Shiba bands for the observation of Majorana modes in atomic chains, \cite{prb_88_020407} the impact of magnetic anisotropy on YSR states, \cite{natcommun_6_8988} the influence of the dimensionality of the superconductor on the spatial coherence of the bound states, \cite{natphys_11_1013} a detailed understanding of the orbital origin of YSR multiplets, \cite{prl_117_186801,natcommun_8_15175} the occurrence of YSR states in proximity-induced superconductor--molecule junctions, \cite{prl_118_117001} the spin polarization of YSR states, \cite{prl_119_197002} and coupling phenomena. \cite{arxiv_30nov17}   

Studies of the effects of charge transport across superconductor -- normal-metal (S--N) and superconductor -- superconductor (S--S) interfaces containing a single magnetic impurity on the bound in-gap states add some degree of complexity since Andreev reflection (AR) \cite{spjetp_19_1228} and Josephson currents \cite{pl_1_251} have to be considered.
Recently, the tunneling process \textit{via} YSR states has been scrutinized for various tunneling currents. \cite{prl_115_087001}
At elevated tunneling rates single-electron and Andreev tunneling have been demonstrated to contribute to the current across the junction and to cause an inversion of the YSR peak height asymmetry.
However, while $\text{d}I/\text{d}V$ spectroscopy of YSR states induced by single magnetic molecules or atoms has thoroughly been investigated in the tunneling range (\textit{vide supra}), the contact range, \textit{i.\,e.}, the ballistic electron transport across the junction in the absence of the tunneling barrier, has remained unexplored so far.
Here, we aim at filling this gap by following the evolution of the excitation spectrum of a superconductor with adsorbed magnetic impurities from the tunneling to the contact range.
Contact to single Mn-Pc molecules adsorbed on Pb(111) is reached for junction conductances exceeding $1\,\text{G}_0$ ($\text{G}_0=2\text{e}^2/\text{h}$: quantum of conductance, h: Planck constant), which is more than an order of magnitude larger than previously reported single-Mn junction conductances. \cite{prl_115_087001}
Using normal-metal W and superconducting Pb tips the evolution of the BCS energy gap together with in-gap YSR states has been analyzed by $\text{d}I/\text{d}V$ spectroscopy and transport calculations for the wide range of junction conductances covering tunneling to contact.
In S--N as well as S--S junctions the spectra unveil a gradual broadening of the YSR peaks with increasing junction conductance and the occurrence of a broad zero-bias resonance close to and at contact.
The calculations trace the broadening of YSR line widths to the increasing tip--molecule hybridization.
For W--Mn-Pc--Pb(111) junctions the zero-bias peak is induced by AR owing to the high transparency of the S--N interface in the contact geometry.
In the case of Pb--Mn-Pc--Pb(111) contacts the zero-bias resonance arises due to a Josephson supercurrent that is carried by phase-dependent Andreev and YSR states.

The article is organized as follows.  
Section \ref{sec:exp} summarizes the experimental results.
In section \ref{sec:theo} our minimal model is presented together with a comparison of simulated data and experimental findings.
Conclusions are drawn in section \ref{sec:con}.

\section{Experiment}
\label{sec:exp}

The experiments were performed with an STM operated at low temperature ($5.5\,\text{K}$) and in ultrahigh vacuum ($10^{-9}\,\text{Pa}$)\@.
Clean and single-crystalline Pb(111) surfaces were obtained after repeated cycles of Ar$^+$ bombardment and annealing.
The cleaned surface was exposed at room temperature to a beam of Mn-Pc molecules sublimated from powder (purity: $99.9\,\%$) in a heated Ta crucible.
Normal-metal tips were fabricated from chemically etched W wire (purity: $99.95\,\%$), while superconducting tips were obtained by coating W tips with Pb substrate material.
The doubling of the BCS energy gap width of Pb(111) was used as an indicator for the superconducting state of the tip.
Constant-height spectra of $\text{d}I/\text{d}V$ were acquired by sinusoidally modulating ($620\,\text{Hz}$) the bias voltage and recording the current response of the junction with a lock-in amplifier.
Modulation voltages of $500\,\mu\text{V}_{\text{pp}}$ and $50\,\mu\text{V}_{\text{pp}}$ were applied in the case of W and Pb-coated W tips, respectively.
All STM images were recorded at constant current with the bias voltage applied to the sample. 

\subsection{Tunneling range}

\begin{figure}
\includegraphics[width=0.95\linewidth]{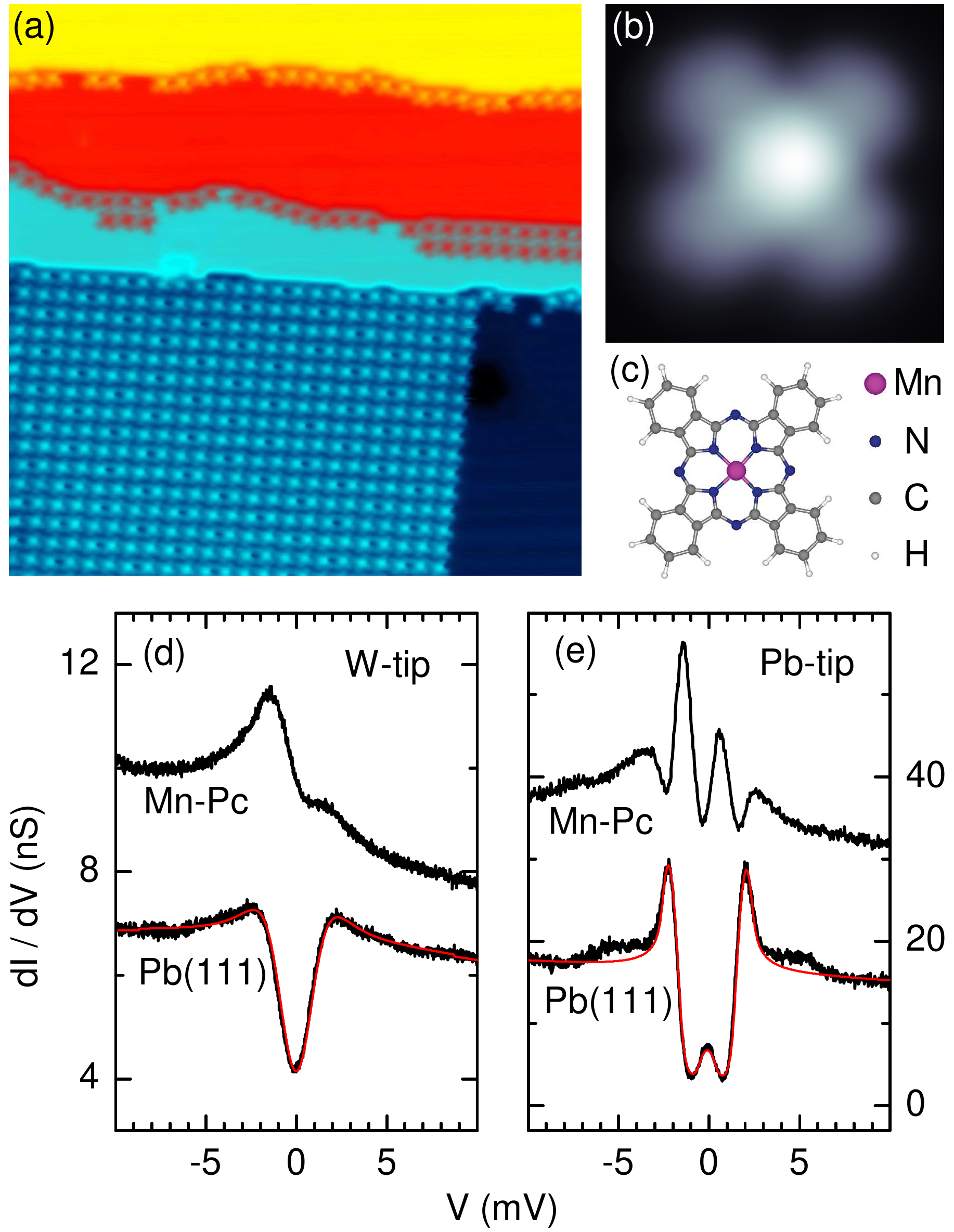}
\caption[fig1]
{(Color online) 
(a) STM image of a Pb(111) surface covered with Mn-Pc molecules in the submonolayer coverage range (tunneling current $I=100\,\text{pA}$, bias voltage $V=100\,\text{mV}$, size: $50\,\text{nm}\times 50\,\text{nm}$)\@.
Step edges are decorated with molecular rows.
Larger substrate terraces exhibit self-assembled ordered arrays with a nearly square superlattice (see text)\@.
(b) STM image of an isolated Mn-Pc molecule on Pb(111) ($100\,\text{pA}$, $-10\,\text{mV}$, $2.6\,\text{nm}\times 2.6\,\text{nm}$)\@.
(c) Stick-and-ball model of Mn-Pc.
(d) Spectra of $\text{d}I/\text{d}V$ acquired atop Pb(111) (bottom) and the center of Mn-Pc (top) with a W tip ($80\,\text{pA}$, $-10\,\text{mV}$)\@.
The spectrum of Mn-Pc is offset by $2\,\text{nS}$\@.  
The solid line is a fit of the BCS quasiparticle DOS to the experimental data.
(e) Like (d), acquired with a Pb-coated W tip ($150\,\text{pA}$, $10\,\text{mV}$)\@.
The spectrum of Mn-Pc is offset by $20\,\text{nS}$\@.
The discussion of the spectral features is presented in the text.}
\label{fig1}
\end{figure}

Figure \ref{fig1}(a) shows an STM image of Mn-Pc molecules adsorbed on Pb(111) in the submonolayer coverage range.
Individual molecules appear with the characteristic crosslike shape and a bright center [Fig.\,\ref{fig1}(b)]\@.
The four lobes are due to the phenyl groups [Fig.\,\ref{fig1}(c)] of the molecule hosting the delocalized $\pi$-electron system.
The Mn atom gives rise to the brightest contrast in STM images due to its $d$-orbitals.
Similar findings were reported for other metallophthalocyanines on metal surfaces. \cite{cpl_392_265,science_309_1542,ss_601_4180,cpl_484_59,jpcc_115_12173,scirep_3_1210,ssr_70_259,langmuir_32_6843}
At this coverage, Mn-Pc molecules decorate step edges and form extended regular
arrays on terraces [lower part of Fig.\,\ref{fig1}(a)]\@.
Since the main focus of this article is the investigation of charge transport the structural properties are briefly summarized.
For details the reader is referred to previous reports, \cite{jcp_134_154703,jpcc_115_21750} which are in agreement with our findings.
The Mn-Pc molecules are assembled in a nearly square arrangement with a nearest-neighbor distance of $1.65\pm 0.02\,\text{nm}$ and with an angle of $89\pm 3^\circ$ enclosed by the lattice vectors of the molecular superstructure.
One of these lattice vectors deviates by $4\pm 1^\circ$ from a Pb(111) crystallographic direction.
The molecular unit cell is not primitive since individual molecules adopt rotational orientations that differ by $8\pm 3^\circ$\@.
As a consequence, the unit cell is a parallelogram containing $4$ molecules.
While the molecular superstructure is incommensurate with the Pb(111) lattice \cite{jcp_134_154703,jpcc_115_21750} isolated Mn-Pc molecules were previously demonstrated to preferably adsorb with their Mn center at on-top sites of the substrate. \cite{jpcc_115_21750}

In a next step, $\text{d}I/\text{d}V$ spectra were acquired atop Pb(111) and Mn-Pc molecules embedded in the array.
Pristine W [Fig.\,\ref{fig1}(d)] and Pb-coated W [Fig.\,\ref{fig1}(e)] tips were used to this end.
The lower spectra in Fig.\,\ref{fig1}(d), (e) show the BCS energy gap of Pb(111)\@.
The increase in energy resolution upon using a superconducting tip is obvious.
The condensation peaks clearly appear at $\approx\pm 2.1\,\text{mV}$ [Fig.\,\ref{fig1}(e)]\@.
In addition, the contribution of thermally excited quasiparticles is visible as a weak zero-bias peak [Fig.\,\ref{fig1}(e)], which is not resolved in the case of a W tip [Fig.\,\ref{fig1}(d)]\@.
Moreover, well resolved features at $\approx\pm 5\,\text{meV}$ are discernible, which are not described by the BCS quasiparticle DOS [solid line in Fig.\,\ref{fig1}(e)]\@.
Their origin is a van Hove singularity in the Pb(111) phonons DOS, \cite{prl_14_108} whose energy has to be corrected for the energy gap. \cite{prl_100_226801} 
The enhancement of the resolution is in accordance with previous reports \cite{prl_100_226801,apl_96_073113} and is due to the sharp onsets of condensation peaks in the tip DOS\@.
Fits of the BCS quasiparticle DOS (solid lines) to the experimental data give rise to gap widths $\Delta_{\text{S}}=1.00\pm 0.30\,\text{meV}$ of the sample and $\Delta_{\text{T}}=0.99\pm 0.30\,\text{meV}$ of the tip.
The extracted $\Delta_{\text{S}}$ corresponds to a sample temperature of $\approx 5\,\text{K}$, which is close to the actual experimental temperature of $5.5\,\text{K}$\@.
The energy gap at $0\,\text{K}$ is $\Delta_0=1.36\,\text{meV}$ with a critical temperature of $T_{\text{c}}=7.2\,\text{K}$\@. \cite{pr_128_591}  
The nearly doubled gap width in $\text{d}I/\text{d}V$ spectra acquired with a Pb-coated W tip demonstrates the superconducting state of the tip.

On top of Mn-Pc molecules, the $\text{d}I/\text{d}V$ spectra differ markedly
due to the presence of intragap YSR states.
Using a normal-metal W tip the energy resolution is not sufficient to observe the individual YSR states.
Rather, a distorted BCS energy gap appears in the $\text{d}I/\text{d}V$ spectra [upper data set of Fig.\,\ref{fig1}(d)]\@.
With a superconducting Pb-coated W tip, however, YSR states can individually be resolved within the gap [upper data set of Fig.\,\ref{fig1}(e)]\@.
Depending on the Mn-Pc molecule embedded in the island the acquired $\text{d}I/\text{d}V$ spectra differ with respect to the position of the YSR-associated peaks and their height ratio.
As exposed above, the molecular array is not commensurate with the Pb(111) lattice, which results in different adsorption sites for Mn-Pc molecules that entail different magnetic couplings of Cooper pairs to the screened magnetic moment of Mn-Pc. \cite{science_332_940}
The asymmetric background visible in all spectra is due to a Kondo resonance, which covers a larger voltage range than the BCS energy gap. \cite{science_332_940}

\subsection{Contact range}

\begin{figure*}
\includegraphics[width=0.75\linewidth]{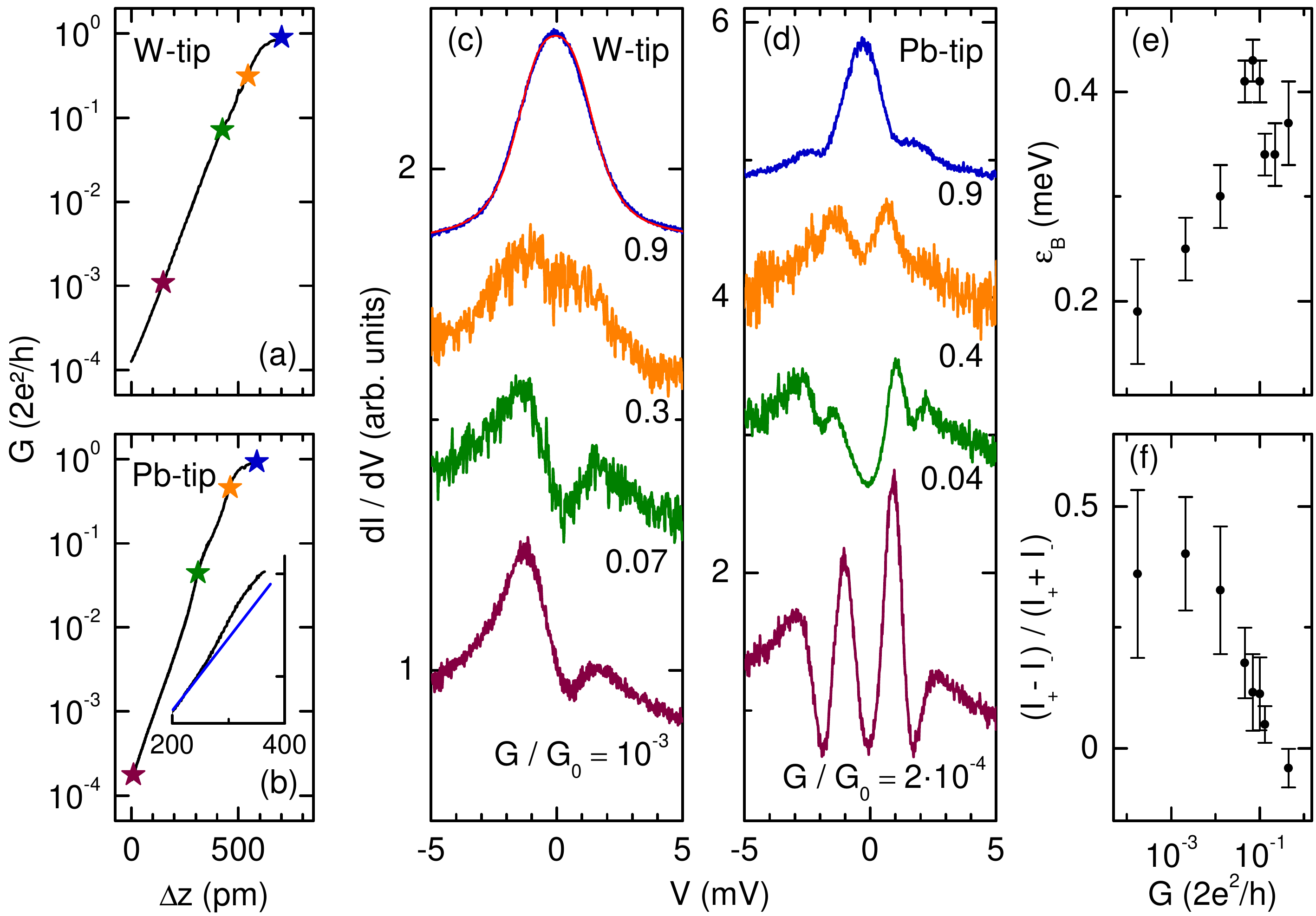}
\caption[fig2]
{(Color online)
Conductance $G$ versus tip displacement $\Delta z$ acquired atop the Mn-Pc center with a W (a) and a Pb-coated W (b) tip.
Stars indicate junction conductances at which $\text{d}I/\text{d}V$ spectra in (c) and (d) were acquired.
$\Delta z=0\,\text{pm}$ is defined by $0.1\,\text{nA}$, $-10\,\text{mV}$ (W tip), $10\,\text{mV}$ (Pb tip)\@.
Inset to (b): Close-up view of the $G$ evolution for $200\,\text{pm}\leq\Delta z\leq 400\,\text{pm}$ showing deviations from the exponential behavior and indicating relaxation effects.
(c), (d) Spectra of $\text{d}I/\text{d}V$ acquired atop the Mn-Pc center with a W (c) and Pb-coated W (d) tip with increasing junction conductance (from bottom to top)\@.
The structured BCS energy gap observed in tunneling spectra evolves into a broad peak centered at zero bias voltage for both junction types.
The solid line in the topmost data set of (c) is a fit of the (broadened) BTK DOS to experimental $\text{d}I/\text{d}V$ data.
The feedback loop was deactivated at (c) $-10\,\text{mV}$, (d) $10\,\text{mV}$ for all spectra. 
Spectra are normalized by $\text{d}I/\text{d}V$ at the feedback loop parameters and vertically offset.
(e) YSR binding energy $\varepsilon_{\text{B}}$ extracted from spectra in (d)\@.
(f) Asymmetry $(I_+ - I_-)/(I_+ + I_-)$ of YSR peak heights extracted from spectra in (d)\@.
}
\label{fig2}
\end{figure*}

We now explore the evolution of the BCS energy gap and YSR spectroscopic signatures with increasing junction conductance.
Some aspects have already been identified for high-conductance junctions in the tunneling range for single Mn atoms adsorbed on Pb(111). \cite{prl_115_087001}
Here, we are particularly interested in the situation at contact and beyond; that is, the electron transport across Mn-Pc will be probed without tunneling barrier between the tip and the molecule.
To this end, pristine and Pb-coated W tips were approached towards an Mn-Pc molecule and the conductance, $G=I/V$, was simultaneously recorded.
Typical $G$ traces are plotted as a function of the tip displacement, $\Delta z$, for a W [Fig.\,\ref{fig2}(a)] and a Pb-coated W [Fig.\,\ref{fig2}(b)] tip.
After exponentially increasing in the tunneling range [$0\,\text{pm}\leq\Delta z\leq 425\,\text{pm}$ (W tip), $0\,\text{pm}\leq\Delta z\leq 375\,\text{pm}$ (Pb tip)] with increasing $\Delta z$, $G$ starts to gradually level off at $\approx 550\,\text{pm}$ for W tips and at $\approx 450\,\text{pm}$ for Pb tips with contact conductances of $\approx 0.9\,\text{G}_0$ (W tip) and $\approx 1\,\text{G}_0$ (Pb tip)\@.
Akin conductance variations were reported for various molecules on diverse surfaces. \cite{jpcm_20_223001,pccp_12_1022}

At the indicated conductances [stars in Fig.\,\ref{fig2}(a), (b)] $\text{d}I/\text{d}V$ spectra were acquired atop the Mn center.
For pristine W tips, a representative evolution is plotted in Fig.\,\ref{fig2}(c)\@.
The asymmetric profile of the spectrum in the tunneling range consists of broad features, which are assigned to the superposition of BCS condensation peaks and YSR signatures riding on a broad Kondo resonance background.
With increasing junction conductance the energy range of the BCS gap is progressively replaced by a zero-bias resonance close to and at contact.

The occurrence of a zero-bias peak for high-conductance junctions is reminiscent of AR\@. \cite{prb_46_5814,prb_79_144522,prl_118_107001} 
For single-C$_{60}$ junctions on Nb(110) it was shown that with a normal-metal W tip the BCS gap evolved into a zero-bias peak with increasing junction conductance up to contact. \cite{prl_118_107001}
On the basis of the Blonder-Tinkham-Klapwijk (BTK) \cite{prb_25_4515} model and transport calculations this evolution was traced to the occurrence of AR at elevated tip--surface coupling. \cite{prl_118_107001}

In the BTK model the transparency of the S--N interface for electron transport is controlled by the Dirac $\delta$ function of weight $Z$, which gives rise to the normal-state transmission $1/(1+Z^2)$\@.
By considering the BTK sample DOS at zero temperature, $\varrho_{\text{BTK}}=\varrho_{\text{N}}(1+A-B)$ [$\varrho_{\text{N}}$: sample DOS in the nonsuperconducting state; $A$ ($B$): energy-dependent probability for Andreev (ordinary) reflection], and taking spectroscopic broadening due to temperature and bias voltage modulation appropriately into account \cite{prl_118_107001} the experimental $\text{d}I/\text{d}V$ data at elevated S--N junction conductance were reproduced [solid line in Fig.\,\ref{fig2}(c)]\@.
The suggested AR scenario is further corroborated by a previous publication reporting the electron transport across S--S single-Mn junctions. \cite{prl_115_087001}
In the range of weak tip--surface coupling the electron transport in tunneling junctions was rationalized in terms of single-particle tunneling; that is, electrons or holes injected into the YSR relax to the quasiparticle continuum.
With increasing injection rate this single-particle picture is no longer applicable.
Rather than relaxing into the quasiparticle continuum the electron injected into the unoccupied YSR state is retroreflected as a hole into the occupied YSR state and propagates as a Cooper pair in the superconducting substrate. \cite{prl_115_087001}

The experimental $\text{d}I/\text{d}V$ spectrum at contact [topmost spectrum in Fig.\,\ref{fig2}(c)] together with the BTK fit (solid line) show that the maximum at $0\,\text{mV}$ exceeds the normal-state $\text{d}I/\text{d}V$ tails ($|V|\geq 10\,\text{mV}$) by a factor 1.5 rather than $2$, which would be expected for ideal AR\@. \cite{prb_25_4515}
The most likely reason for this observation is the temperature-induced broadening of the AR peak.
A reduction of the ideal zero-bias AR conductance may likewise occur in the case of a spin-polarized current across the junction, which owing to the presence of a magnetic molecule may be expected.
Indeed, S--N quantum point contacts were suggested and used to extract the spin polarization \textit{via} the reduced AR conductance \cite{prl_74_1657,science_282_85,prl_81_3247,prl_86_5585,prb_65_020508,prb_66_212403} on the basis of a modified BTK model. \cite{science_282_85,prb_63_104510} 
In the present case, however, the spectroscopic broadening due to the finite temperature exceeds $30\,\%$ of the superconducting order parameter, which hampers the univocal assignment of the AR conductance decrease to spin polarization of the current.

Using Pb-coated W tips more spectroscopic details are available owing to the superconducting state of the tip and the concomitant increase of resolution.
Figure \ref{fig2}(d) shows the evolution of the BCS energy gap with clearly resolved BCS condensation and YSR peaks from tunneling (bottom) to contact (top) ranges.
The most obvious change is the filling of the BCS gap with a single zero-bias resonance close to and at contact.
At contact [topmost data in Fig.\,\ref{fig2}(d)] the maximum of the resonance exceeds the normal-state $\text{d}I/\text{d}V$ at $|V|\geq 10\,\text{mV}$ by $\approx 2.3$\@.
These observations are reminiscent of Cooper pair tunneling observed from symmetric \cite{prl_87_097004} and asymmetric \cite{prb_74_132501} STM S--S junctions.
Indeed, the calculations discussed in the following section show that this zero-bias peak results from a Josephson supercurrent that is carried by phase-averaged Andreev and YSR levels.
In the range of junction conductances ($G\leq 0.4\,\text{G}_0$) where the YSR features can be resolved as individual peaks their energy $|\varepsilon_\text{B}|$ slightly increases [Fig.\,\ref{fig2}(e)], while the asymmetry of peak heights, $(I_+-I_-)/(I_++I_-)$ [$I_\pm$: signal strength of the YSR state in $\text{d}I/\text{d}V$ spectra at positive ($+$) and negative ($-$) bias voltage, Fig.\,\ref{fig2}(f)], is reduced with increasing conductance and even crosses $0$ close to contact.
YSR energies and peak height asymmetries were extracted from fits to the experimental data where the BCS quasiparticle DOS of the tip was convoluted with the sample DOS represented by two Lorentzians for the YSR states and a symmetric gap around the Fermi energy. \cite{science_332_940}
The observed increase of $|\varepsilon_\text{B}|$ with tip approach may be interpreted in terms of a reduced magnetic exchange interaction of the central Mn atom and the Pb(111) Cooper pairs in the following sense.
The conductance-versus-displacement data depicted in the inset to Fig.\,\ref{fig2}(b) show deviations from a simple exponential behavior in the range $0.01\,\text{G}_0\lesssim G\lesssim 0.5\,\text{G}_0$\@.
Such behavior is indicative of relaxations in the junction geometry. \cite{prl_94_126102,prl_98_065502,njp_9_153,nl_8_1291,njp_11_125006,prl_104_176802,prl_105_236101,nl_11_3593,prb_86_180406,prb_93_235402,prl_120_076802,prb_97_075418}
In the present case the  Mn atom may be attracted towards the tip apex.
Concomitantly, Mn is lifted away from the surface and effectively reduces the exchange interaction.
The trading of peak height is in agreement with findings for single Mn atoms on Pb(111)\@. \cite{prl_115_087001}

\section{Theory}
\label{sec:theo}

\begin{figure}
\includegraphics[width=0.9\columnwidth]{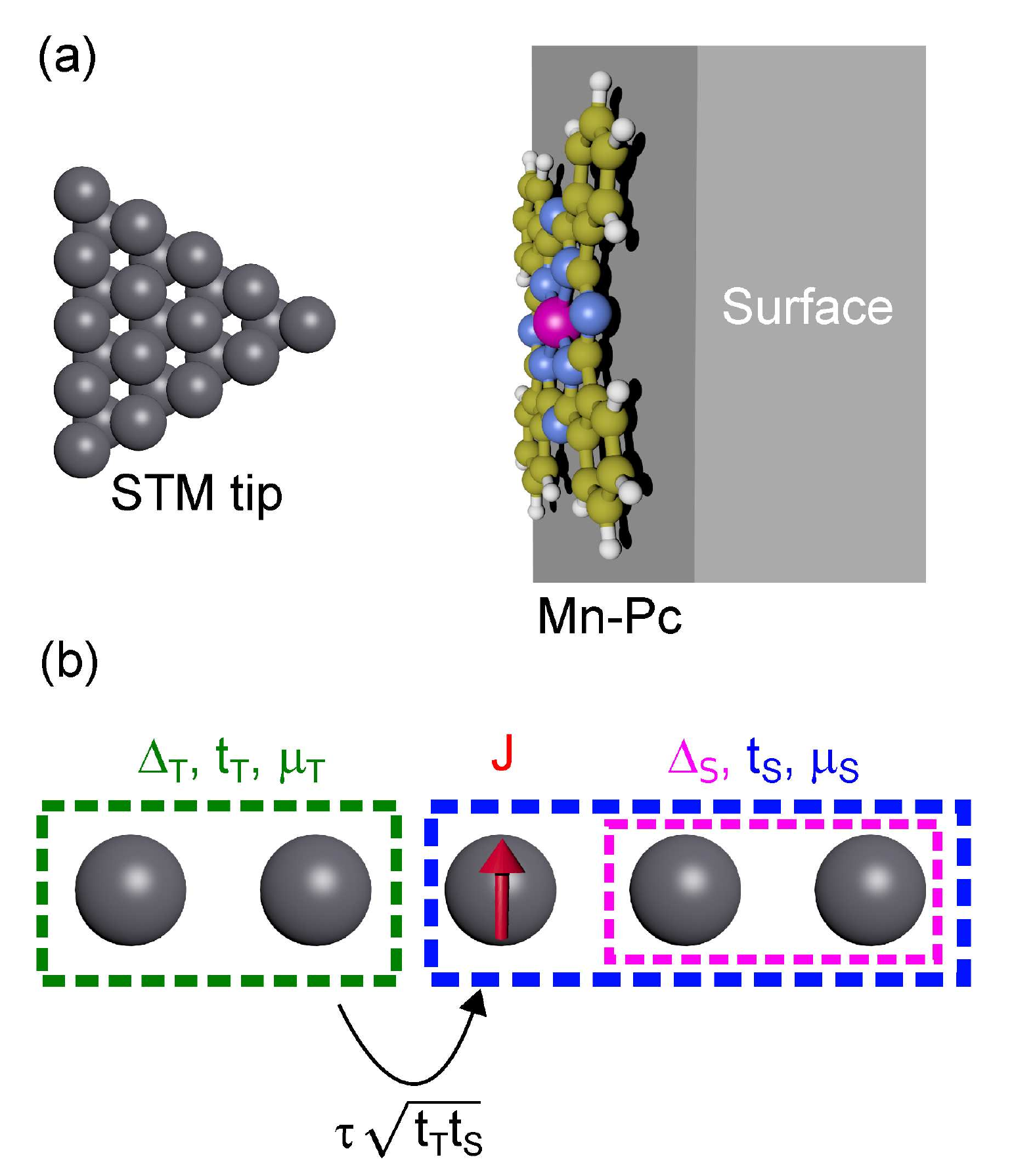}
\caption[fig3]
{(Color online)
(a) Sketch of the experimental geometry showing an STM tip, a deposited Mn-Pc molecule and the superconducting substrate. 
(b) Sketch of the minimal model employed for the geometry in (a)\@.
The tip (green dashed line) and the superconducting substrate (purple) are modeled as semi-infinite linear atomic chains with superconducting order parameters $\Delta_\text{T}$ and $\Delta_\text{S}$, respectively.
Electrons in the tip (substrate) move in a single band with hopping parameter
$t_\text{T}$ ($t_\text{S}$) and chemical potential $\mu_\text{T}$
($\mu_\text{S}$)\@.
At the molecule site, a local magnetic exchange interaction, $J$, is introduced.
The hopping amplitude $\tau\sqrt{t_\text{T}t_\text{S}}$ controls the tip--molecule--substrate coupling with $0\leq\tau\leq 1$ the total transparency of the junction.
}
\label{fig3}
\end{figure}

\subsection{Hamiltonian and methods}

In this section, a Hamiltonian is proposed in order to capture the physics of
the experimental findings, including the crossover from the tunneling to the
contact range. 
The experimental setup is schematically sketched in Fig.\,\ref{fig3}(a)\@.
The minimal model reduces the complex setup to two semi-infinite linear atomic chains representing the tip [green dashed line in Fig.\,\ref{fig3}(b)] and the molecule-covered substrate (blue dashed line)\@.
The Hamiltonian  describes the electronic states in the tip, $\mathcal{H}_{\text{T}}$, the superconducting substrate and the magnetic molecule, ${\cal H}_{\text{SM}}$, and the coupling, $\mathcal{V}$, between them, \textit{i.\,e.}, 
\begin{equation}
\mathcal{H}=\mathcal{H}_{\text{T}}+\mathcal{H}_{\text{SM}}+\mathcal{V}.
\label{eq1}
\end{equation}

The Hamiltonian describing the substrate and the molecule reads
\begin{eqnarray}
\mathcal{H}_{\text{SM}}&=&\sum_{i\geq 0,\sigma}\left[t_\text{S}\left(c^\dagger_{i,\sigma}c_{i+1,\sigma}+c^\dagger_{i+1,\sigma}c_{i,\sigma}\right)+\mu_S c^\dagger_{i,\sigma}c_{i,\sigma}\right] \nonumber \\
&+&J \left(c^\dagger_{0,\uparrow}c_{0,\uparrow}-c^\dagger_{0,\downarrow}c_{0,\downarrow}\right) \nonumber \\
&+&\sum_{i>0}\Delta_\text{S}\left(c_{i,\uparrow}c_{i,\downarrow}+\text{h.c.}\right),
\label{eq2}
\end{eqnarray}
where $c_{i,\sigma}^\dagger$ ($c_{i,\sigma}$) denotes the creation (annihilation) of an electron in the substrate ($i>0$) or at the molecule ($i=0$) with spin $\sigma\in\{\uparrow,\downarrow\}$ (arrows indicate the spin direction with respect to the magnetic moment of the molecule)\@.
The electrons are assumed to move in a single band with hopping $t_\text{S}$, an energy offset $\mu_\text{S}$ and uniform $s$-wave superconducting pairing $\Delta_\text{S}$ vanishing at the molecule ($i=0$)\@.
At the site of the molecule an exchange interaction, $J$, is added [Fig.\,\ref{fig3}(b)]\@. 

Analogously, for the tip ($i<0$) a single band with hopping $t_\text{T}$,
energy offset $\mu_\text{T}$ and superconducting $s$-wave order parameter
$\Delta_\text{T}$ is used; that is,
\begin{eqnarray}
\mathcal{H}_{\text{T}}&=&\sum_{i<0,\sigma}\left[t_\text{T}\left(c^\dagger_{i,\sigma}c_{i+1,\sigma}+c^\dagger_{i+1,\sigma}c_{i,\sigma}\right)+\mu_\text{T} c^\dagger_{i,\sigma}c_{i,\sigma}\right] \nonumber \\
&+&\sum_{i<0} \Delta_\text{T}\left(c_{i,\uparrow}c_{i,\downarrow}+c^\dagger_{i,\downarrow}c^\dagger_{i,\uparrow}\right).
\label{eq3}
\end{eqnarray}

The coupling between the tip and the molecule-covered substrate is mediated \textit{via} the molecule at site $i=0$ with a hopping amplitude $\tau\sqrt{t_\text{T}t_\text{S}}$, where the dimensionless parameter $\tau$ controls the total transparency of the junction, \textit{i.\,e.},
\begin{equation}
\mathcal{V}=\tau\sqrt{t_\text{T}t_\text{S}}\sum_\sigma\left(c^\dagger_{-1,\sigma}c_{0,\sigma}+c^\dagger_{0,\sigma}c_{-1,\sigma}\right).
\label{eq4}
\end{equation}
In this way, the tunneling regime corresponds to $\tau\ll 1$, whereas $\tau=1$ mimics the contact range with full transparency.

Choosing $\Delta_\text{T}=0$ ($\Delta_\text{T}\neq 0$) a normal-metal (superconducting) tip may be modeled.
In addition, the variation of $\tau$ from $\tau\ll 1$ to $\tau\simeq 1$ describes the transition from the tunneling to the contact range. 
In general, the effective hopping of the tip $t_\text{T}$ and the substrate $t_\text{S}$, and their energy offsets $\mu_\text{T}$ and $\mu_\text{S}$ are expected to be different.
For the sake of simplicity, however, we choose $t_\text{T}=t_\text{S}=t$ and $\mu_\text{T}=\mu_\text{S}=\mu$.
This assumption does not change the results qualitatively. 

\begin{figure}
\includegraphics[width=\linewidth]{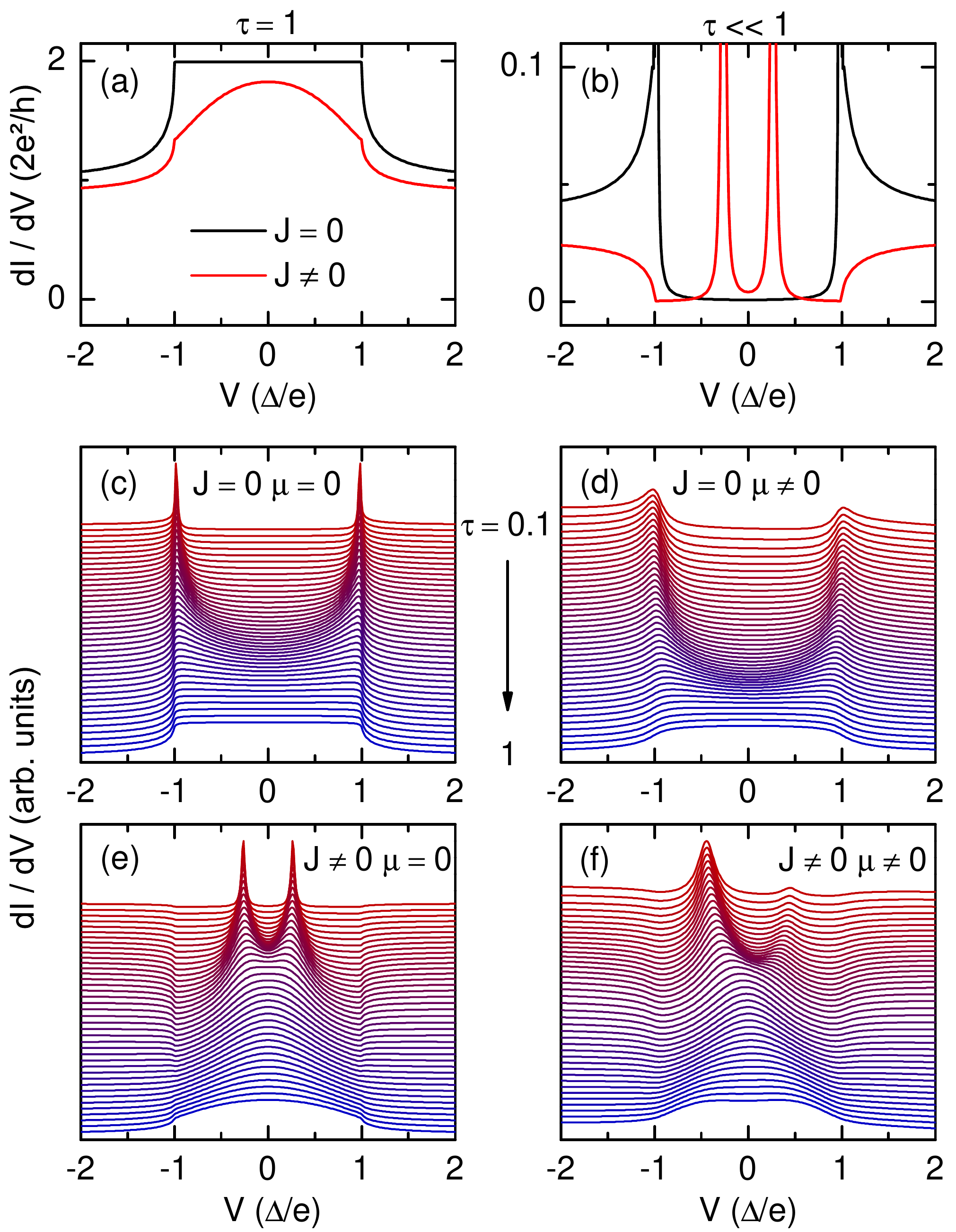}
\caption[fig4]
{(Color online) 
Calculated $\text{d}I/\text{d}V$ in the limit of a perfect ($\tau=1$) contact (a) and in the tunneling ($\tau\ll 1$) range (b), without ($J=0$) and with ($J\neq 0$) magnetic impurity.
Evolution of $\text{d}I/\text{d}V$ without (c), (d) and with (e), (f) a magnetic impurity as a function of the transparency $\tau$.
The data of panels (a)--(c), (e) were calculated for the electron-hole symmetric case ($\mu=0$) and with negligible broadening of the spectral function ($\delta\ll\Delta$)\@. 
The data of panels (d), (f) were calculated for $\mu\neq 0$ and $\delta = 0.1\,\Delta$\@. 
$\text{d}I/\text{d}V$ data in (c)--(f) are normalized to their maximum value to clearly see the qualitative changes with increasing transmission (from top to bottom)\@.
}
\label{fig4}
\end{figure}

The previous Hamiltonian defines two semi-infinite systems [Fig.\,\ref{fig3}(b)]\@. 
In the case of the normal-metal tip $\text{d}I/\text{d}V$ is calculated by means of the $S$-matrix formalism. 
The spectral function $g_\text{X}$ ($\text{X}=\text{T},\text{S}$) of the tip and the surface is derived by solving the Dyson equation $g_\text{X} = (\varepsilon + \text{i}\delta - H_\text{X} - \gamma_\text{X}^\dagger g_\text{X}\gamma_\text{X})^{-1}$, where $\varepsilon=\text{e}V$ is the electron energy, $\delta$ a damping constant, $H_\text{X}$ the on-site matrix, and $\gamma_\text{X}$ the hopping matrix for the specific part. 
These two Green functions are the basis for computing the full Green function of a subsystem involving part of the tip and part of the substrate surface as $g=(\varepsilon + \text{i}\delta - h - \Sigma_\text{T} - \Sigma_\text{S})^{-1}$, where $\Sigma_\text{X} = \gamma_\text{X}^\dagger g_\text{X} \gamma_\text{X}$ is the induced self-energy and $h$ the on-site Hamiltonian of the subsystem.
With the full Green function $g$ and the hybridization functions $\Gamma_\text{X} = \text{i}(\Sigma_\text{X}- \Sigma_\text{X}^\dagger)$ the $S$-matrix follows from the Fisher-Lee approach. \cite{prb_23_6851}
Taking the $S$-matrix as $S = \begin{pmatrix} \rho_\text{T} & \tau_\text{S} \\ \tau_\text{T} & \rho_\text{S} \end{pmatrix}$, the different transmission ($\tau_\text{X}$) and reflection ($\rho_\text{X}$) matrices can be calculated as $\rho_\text{X} = -\mathbf{1} + \text{i}\sqrt{\Gamma_\text{X}}g_{\text{XX}}\sqrt{\Gamma_\text{X}}$ and $\tau_\text{X} = \text{i}\sqrt{\Gamma_\text{X}}g_{\text{X}\bar{\text{X}}}\sqrt{\Gamma_{\bar{\text{X}}}}$ with $\mathbf{1}$ the unity matrix.
Focusing on an incoming electron from the normal-metal tip, the total conductance reads $G = N - R_{\text{ee}} + R_{\text{eh}}$, \cite{prb_25_4515} where $N$ is the number of channels, $R_{\text{ee}}$ the electron-electron  and $R_{\text{eh}}$ the electron-hole reflection coefficient.
$R_{\text{ee}}$ and $R_{\text{eh}}$ are calculated as $R=\text{Tr}(\bar\rho\bar\rho^\dagger)$, where $\bar\rho$ is the relevant block (electron-electron or electron-hole) of the reflection matrix.

This procedure is only valid if one of the leads is a normal metal, which allows to define electron reflection and transmission coefficients. 
In the case of two superconducting leads the calculation of the current is more complicated and we will adopt a different strategy (\textit{vide infra}) to study the spectral function and its dependence on the superconductor phase difference.   

\subsection{Normal-metal tip}

In this section the S--N junction is considered, \textit{i.\,e.}, $\Delta_\text{S}=\Delta$, $\Delta_\text{T}=0$, in the idealized situation where $\mu=0$ and the spectral broadening vanishes.
In this case $\text{d}I/\text{d}V$ can be calculated within the BTK theory using the scattering-matrix formalism.
Figure \ref{fig4}(a) shows the calculated $\text{d}I/\text{d}V$ for the perfectly transparent ($\tau=1$) S--N interface without (top graph) and with (bottom) magnetic impurity.
For $J=0$ the AR reflection plateau is obtained in the voltage range $|V|\leq\Delta/\text{e}$. 
At finite exchange interaction ($J\neq 0$) the AR plateau transforms into a broad resonance peak due to backscattering effects.
Therefore, the occurrence of the broad $\text{d}I/\text{d}V$ resonance in the experiments can be interpreted as the onset of AR\@.

In the tunneling range [$\tau\ll 1$, Fig.\,\ref{fig4}(b)] the calculated $\text{d}I/\text{d}V$ reflects the spectral function of the surface.
Without magnetic impurity  $(J=0)$ the spectral function consists of the BCS energy gap, whereas in the presence of the magnetic impurity ($J\neq 0$) YSR states appear in the gap.
In Fig.\,\ref{fig4}(c)--(f) the evolution of $\text{d}I/\text{d}V$ with increasing $\tau$ (top: $\tau=0.1$, bottom: $\tau=1$) is shown.
For $J=0$ [Fig.\,\ref{fig4}(c), (d)] our method correctly reproduces the standard evolution from the tunneling range, where $\text{d}I/\text{d}V$ is proportional to the BCS DOS of the substrate, to the AR regime.
While for $\mu=0$ [Fig.\,\ref{fig4}(c)] the calculated data are symmetric with respect to zero voltage, using $\mu\neq 0$ [Fig.\,\ref{fig4}(d)] removes the electron-hole symmetry and results in asymmetric $\text{d}I/\text{d}V$ curves. 
For $J\neq 0$ [Fig.\,\ref{fig4}(e), (f)] the $\text{d}I/\text{d}V$ data exhibit in-gap YSR peaks for small $\tau$ that gradually evolve into a broad zero-voltage resonance with increasing $\tau$.  
In particular, the case without electron-hole symmetry [Fig.\,\ref{fig4}(f)] together with some sizable broadening matches the experimental situation of Fig.\,\ref{fig2}(c) quite well.
It captures the different signal strengths of the YSR peaks in the tunneling range as well as their evolution into a single peak with increasing $\tau$.
Therefore, the experimentally observed zero-bias peak close to and at contact reflects a superposition of the AR signature and broadened YSR levels. 

\begin{figure}
\includegraphics[width=\columnwidth]{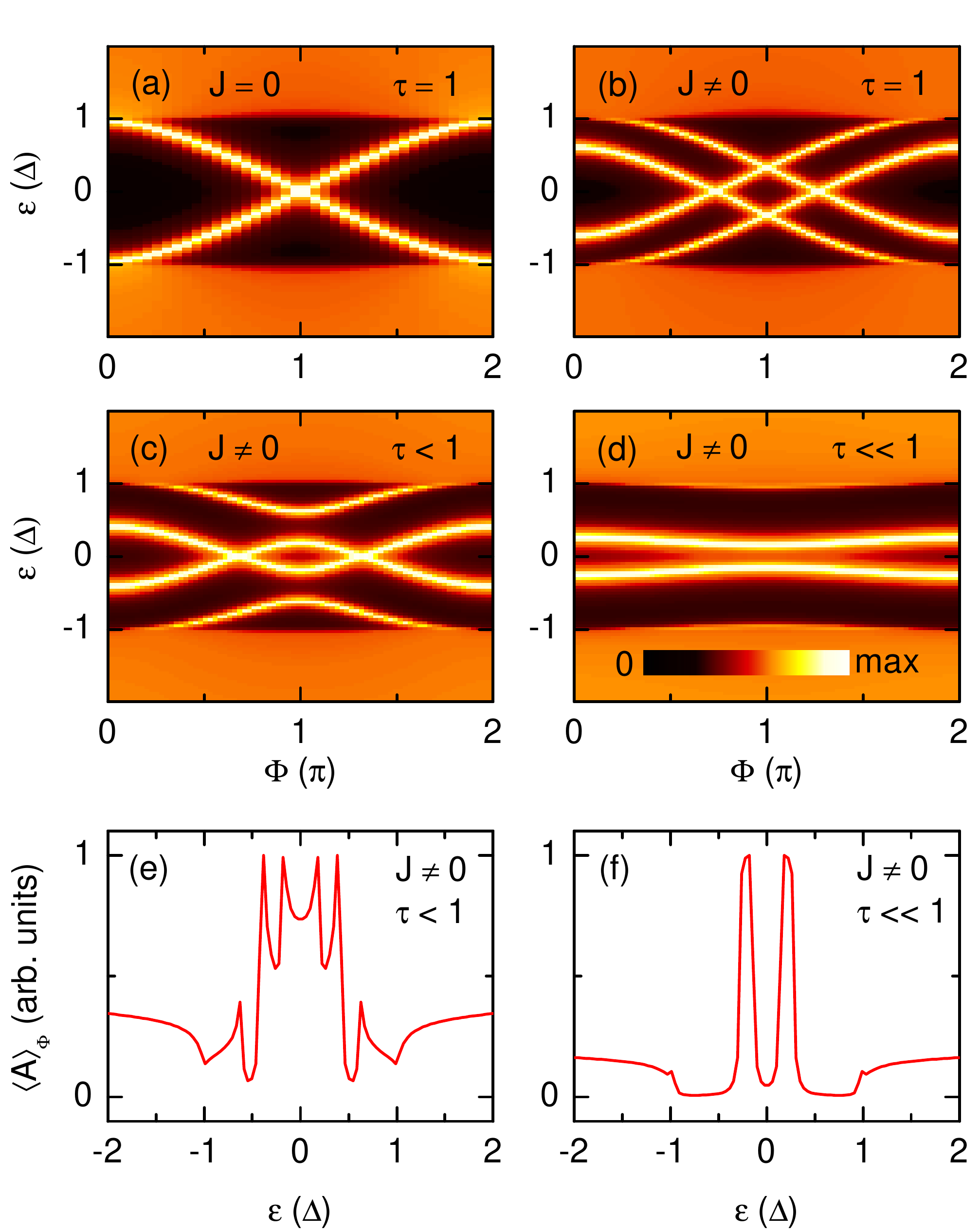}
\caption[fig5]
{(Color online)
Contour plots of the spectral function of an S--S junction depending on the phase difference $\Phi$ between the superconducting electrodes.
(a) $J=0$, $\tau=1$, evolution of the Andreev bound state, which for $\Phi=0, 2\pi$ is located at the edges of the BCS energy gap and evolves into an in-gap state for $0<\Phi<2\pi$.
(b) $J\neq 0$, $\tau=1$, occurrence of in-gap YSR states.
(c) $J\neq 0$, $\tau<1$, the YSR branches become decoupled from the continuum of quasiparticle states for not perfectly transparent junctions.
(d) $J\neq 0$, $\tau\ll 1$, the YSR energy is essentially independent of the phase difference for tunneling junctions.
(e) and (f) Phase-averaged spectral function, $\langle A\rangle_\Phi$, of (c) and (d), respectively.
$\langle A\rangle_\Phi$ may be interpreted as the average spectral function for an electron that enters into the junction at some arbitrary time.  
}
\label{fig5}
\end{figure}

\subsection{Superconducting tip}

The situation of a Pb tip in the experiments can be accounted for by $\Delta_\text{T}\neq 0$\@. 
Due to the Josephson effect \cite{pl_1_251} the finite bias voltage applied across the S--S junction leads to a time-dependent relative phase between the two superconductors. \cite{becu_1984,njp_15_075019}
For the sake of simplicity, we set $\Delta_\text{S}=\Delta$ and $\Delta_\text{T }=\Delta\,\text{e}^{\text{i}\Phi}$ with $\Phi$ the time-dependent phase difference between the substrate and the tip.
The transition between the tunneling and the contact range can qualitatively be understood by comparing the spectral function of the junction in the different transport ranges for the entire range of $\Phi$, \textit{i.\,e.}, $A(\Phi,\varepsilon)=-\text{Im} [g(\Phi,\varepsilon)]/\pi$.
The full Green function $g(\Phi,\varepsilon)$ is calculated by means of the Dyson equation
$g(\Phi,\varepsilon)=[\varepsilon+\text{i}\delta-h-\Sigma_\text{T}(\Phi,\varepsilon)-\Sigma_\text{S}(\varepsilon)]^{-1}$,
which explicitly shows the dependence on the superconducting phase difference.
In the following, the spectral function represents the sum over electron and hole sectors and therefore appears symmetric for positive and negative energies, respectively.
An asymmetry is present if electron and hole sectors are considered separately. 

In the contact range, $\tau=1$, and without magnetic impurity, $J=0$, the spectra of $A$ of the interface show branches at positive and negative energies crossing the BCS energy gap as $\Phi$ increases from $0$ to $2\pi$ [Fig.\,\ref{fig5}(a)]\@.
These branches correspond to discrete Andreev levels that carry Cooper pairs from one superconductor to the other and, thus, signal the presence of a finite Josephson supercurrent.
Similar conclusions were drawn from findings obtained for microscale nanotube--superconductor junctions. \cite{natphys_6_965}  
The slope, $\text{d}\varepsilon/\text{d}\Phi$, is a measure of the magnitude of
the Josephson current.
Upon taking the magnetic impurity into account ($J\neq 0$), the original Andreev branch splits into two for positive and negative energies as shown in Fig.\,\ref{fig5}(b)\@. 
These two branches become disconnected from the continuum of quasiparticle states when the transparency of the junction is reduced [Fig.\,\ref{fig5}(c)]\@.
In the tunneling range, $\tau\ll 1$, the original two branches become an almost flat band signaling the presence of the conventional YSR states of the substrate [Fig.\,\ref{fig5}(d)]\@.
The Andreev levels have essentially vanished in this transmission range.

At finite bias voltage, the phase difference evolves in time. 
Therefore, the phase-averaged spectral function, $\langle A\rangle_\Phi$, was calculated [Fig.\,\ref{fig5}(e), (f)]\@.  
Close to contact, $\tau<1$, $\langle A\rangle_\Phi$ exhibits a series of peaks around zero energy within the BCS gap [Fig.\,\ref{fig5}(e)]\@.  
Additional broadening, which may be caused by a finite temperature, would lead to  a broad  resonance centered around zero energy.
In comparison, in the tunneling range, $\tau\ll 1$, only two sharp resonances appear inside the BCS gap [Fig.\,\ref{fig5}(f)]\@.
These findings for the phase-averaged spectral function resembles the experimentally observed transition from the two in-gap peaks to a broad in-gap resonance close to and at the contact range.
Before concluding, we remark that the spectral function allows to determine whether bound states exist within the superconducting energy gap and to describe their evolution with increasing junction conductance.
However, the spectral function is not directly comparable with the experimental $\text{d}I/\text{d}V$ data.
Indeed, the gapped energy region of the spectral function ranges from $-\Delta$ to $\Delta$, while the $\text{d}I/\text{d}V$ spectra exhibit the doubled BCS gap extending from $-2\Delta$ to $2\Delta$\@.
Only more sophisticated transport calculations would improve the simulations.

\section{Conclusion}  
\label{sec:con}

Charge transport across a single magnetic molecule has been experimentally and theoretically studied for S--N and S--S STM junctions from tunneling to contact ranges.
While in the tunneling range electron transport is mainly mediated by YSR states, the AR regime becomes increasingly important for larger junction conductance.
In both types of junctions the conductance increase gives rise to the gradual evolution of a zero-energy resonance that reflects the broadening of YSR states due to the tip--molecule hybridization and the onset of AR\@.
Cooper-pair transport in highly conductive S--S junctions is carried by phase-dependent Andreev and YSR levels that cause a finite Josephson supercurrent.
The experimental approach of probing $I$--$V$ characteristics of single-molecule S--N and S--S junctions from tunneling to contact ranges enables the unambiguous identification of YSR and AR states and their discrimination from other zero-bias features, such as Majorana zero modes.

\acknowledgments
Financial support by the Deutsche Forschungsgemeinschaft through Grant No.\,KR 2912/10-1, the FCT (project PTDC/FIS-NAN/4662/2014 and P2020-PTDC/FIS-NAN/3668/2014), and the MINECO-Spain (MAT2016-78625-C2) is acknowledged.
J.\,L.\,L.\, acknowledges financial support from the ETH Fellowship program.

\bibliographystyle{apsrev4-1}
%

\end{document}